\documentclass[11pt]{article} % use larger type; default would be 10pt

\usepackage[utf8]{inputenc} % set input encoding (not needed with XeLaTeX)

\usepackage{geometry} % to change the page dimensions
\usepackage{graphicx} % support the \includegraphics command and options

\usepackage{booktabs} % for much better looking tables
\usepackage{array} % for better arrays (eg matrices) in maths
\usepackage{paralist} % very flexible & customisable lists (eg. enumerate/itemize, etc.)
\usepackage{verbatim} % adds environment for commenting out blocks of text & for better verbatim
\usepackage{subfig} % make it possible to include more than one captioned figure/table in a single float
\usepackage{fancyhdr} % This should be set AFTER setting up the page geometry
\usepackage{sectsty}
\allsectionsfont{\sffamily\mdseries\upshape} % (See the fntguide.pdf for font help)
\usepackage[nottoc,notlof,notlot]{tocbibind} % Put the bibliography in the ToC
\usepackage[titles,subfigure]{tocloft} % Alter the style of the Table of Contents

\title{Evolution of statistical averages: an interdisciplinary proposal using the Chapman-Enskog method  }

\author{A. Mariscal-Sanchez, A. Sandoval-Villalbazo \\
 Depto. de Fisica y Matematicas, Universidad Iberoamericana,\\
Prolongacion Paseo de la Reforma 880, Mexico D. F. 01219, Mexico.
}
\date{January 2015}

\begin{document}

\maketitle

\begin{abstract}
This work examines the idea of applying the Chapman-Enskog (CE) method for approximating the solution of the Boltzmann equation beyond the realm of physics, using an information theory approach. 
Equations describing the evolution of averages and their fluctuations in a generalized phase space are established up to first order in the Knudsen parameter, which is defined as the ratio of the time between interactions (mean free time) and a characteristic macroscopic time. 
Although the general equations here obtained may be applied in a wide range of disciplines, in this paper only a particular case related to the evolution of averages in speculative markets is examined.

\end{abstract}
\bigskip
\bigskip

\newpage

\newpage

\section{Introduction}

Establishing analogies between physical and other everyday life situations has proven to be advantageous. The depth and accuracy provided by statistical physics and, particularly, kinetic theory can be used to model the evolution of  relevant features involving a large amount of agents. 
Being able to model price distributions could be helpful in the prediction of speculative market turbulence and could potentially be used to avoid economic stock market crashes. Historically, economical crises are known to have a great impact on our society; it is now most important to devise new ways to model and possibly prevent such events. One appropriate method to accomplish this task is to apply the Boltzmann equation onto these matters. 
Kinetic theory can be used to describe the evolution of phenomena such as flocking behavior and speculative markets. Because of its multi-agent nature, kinetic theory is apt to describe such systems that also have multiple interacting variables. Moreover, the relationship between the time among interactions (in finance, the time taken by a stock to change its price) and the characteristic macroscopic times (trading hours) yields a  suitable Knudsen number which will characterize the appropriate hydrodynamic regime.

\bigskip
Information theory proposes that statistical mechanics can be considered as an application of the broader formalism of mathematical statistics \cite{Jaynes} \cite{Shannon}. Accordingly, most methods of kinetic theory which rely on a maximum (information) entropy principle are subject to be non-exclusively applied to physics. In fact, the basic ideas of \textit{econophysics} are based on  probabilistic and statistical methods taken from statistical physics. 
 In both information theory and statistical mechanics the quantity  $H=-\sum p_{i} ln p_{i}$ represents an entropy that upon maximization allows the establishment of a least-biased distribution function. In this sense, the concept of entropy is not limited to the scope of physics. 
\\
\\ Classical kinetic theory leads to a set of partial equations  that describe the evolution of averaged quantities corresponding to the main properties of a given physical microscopic system; these are called \emph{transport equations}. The starting point is a master equation (i.e. the Boltzmann equation) that describes the state variables through the evolution of the distribution function of single particles. When the Boltzmann equation is multiplied by the collisional invariant $\psi$ and integrated in the velocity space,  the resulting equation is called  \textit{Enskog transport equation}.
The distribution function allows to compute averages at different regimes in terms of an order parameter which in hydrodynamics is called the Knudsen number  $\varepsilon$ \cite{cc} .  This parameter is given by the proportion between the mean free time and the characteristic time, thus the distribution function can be written in terms of powers of $\varepsilon$.  The \textit{Euler regime} corresponds to averages taken with respect to the equilibrium distribution function, while the \textit{Navier-Stokes} transport equations are obtained at order $\varepsilon$.
\bigskip
\\ In 1954, Bhatnagar, Gross, and Krook \cite{BGK} proposed a simplification to the right hand side of the Boltzmann equation now known as the BGK kernel  approximation. The BGK model is the basic tool to be used in this paper. Despite
its simplicity, it is able to describe the evolution of a set of particles towards their equilibrium state predicting useful, verifiable results. 
This type of kernel is not restricted to collisional processes, its form predicts an evolution to the equilibrium distribution function in a characteristic time $\tau$. This suggests that the method of establishing transport equations in terms of powers of a Knudsen number does not depend on physical or mechanical laws but in maximum information entropy and stochastic processes considerations. Therefore the mechanism appears to be appropriate to be applied in any situation describable by averages over distribution functions taken out of equilibrium in terms of powers of the Knudsen parameter.  Hence it is interesting to establish a generic set of transport equations, both in its non-linear and its linearized forms, in order to apply the vast range of results of hydrodynamics to a large set of interdisciplinary situations such as financial markets and biophysics.
\\
\\
To accomplish this task we have divided this work as follows: in section two a solution to the one-dimensional Boltzmann equation is obtained up to first order in the Knudsen parameter using the Chapman-Enskog method. The transport equations are established in this regime, and  corresponding transport coefficients are established in terms of characteristic time $\tau$, local equilibrium variables and Boltzmann's constant $k$. This section is included in order to make the paper self-contained, so that the reader can appreciate the limits of the analogies here proposed.  In section three, the generic set of linearized  transport equations for an interdisciplinary case is written in terms of three basic state variables not necessarily related to hydrodynamics. It is compelling to notice that in this representation, there is no need to identify physical-type parameters such as the atomic mass or Boltzmann's constant. Finally, section four is devoted to a discussion of the formalism proposed in this work.

\section{ Navier-Stokes transport equations in a physical system (simple, 1- dimension dilute gas)}

\subsection{ The Boltzmann equation and equilibrium distribution.}

Most of the material comprised in this section refers to a particular case of a general approach to kinetic theory included in classic references of the subject \cite{cc}. We decided to include this material in the present work in order to show how a hydrodynamic-type approach to the evolution of statistical averages can be applied beyond the realm of physical theories. Our starting point is the one-dimensional Boltzmann equation which for a one component dilute gas in the BGK approximation reads:

\begin{equation}
\frac{\partial f}{\partial t}+ v \frac{\partial f}{\partial x}=- \frac{f-f^{(0)}}{\tau}\label{eq:1}
\end{equation}
where $f$ is the distribution function which contains the equilibrium contribution plus the first correction in the Knudsen parameter ($\varepsilon$) :
\begin{equation}
f=f^{(0)}+\varepsilon f^{(1)}\label{eq:o}
\end{equation}
The equilibrium distribution $f^{(0)}$ in Eq. (\ref{eq:1}-\ref{eq:o})  is given by the  Maxwell-Boltzmann function:
\begin{equation}
f^{(0)}=n (\frac{m}{2 \pi k T})^{1/2} e^{\frac{-m (v-u)^{2}}{2 k T}}
\label{eq:mb}
\end{equation}
The averages of the thermodynamic variables are given by:
\begin{equation}
\langle \psi \rangle=\frac{1}{n} \int f \psi dv
\end{equation}
According to this definition, the number of particles per unit of volume reads:
\begin{equation}
\langle n \rangle=\int  f dv
\end{equation}
the  average velocity is given by:
\begin{equation}
\langle v \rangle=\frac{1}{n}\int v f dv=u
\end{equation}
the chaotic velocity is defined as $c=v-u$, and the average energy per particle is:
\begin{equation}
\langle \frac{1}{2}m v^{2} \rangle=\frac{1}{n}\int \frac{1}{2}m v^{2} f dv
\end{equation}
\\
It is important to remark that:
\begin{equation}
\langle  c \rangle=0
\end{equation}
and
\begin{equation}
\langle  c^{2} \rangle=\frac{1}{n} \int c^{2} f^{(0)} dv=\frac {kT}{m}
\end{equation}
\\
Eq. (9) allows to relate temperature  with the mean molecular energy of the system.

\subsection{ Enskog's transport equation.}

Multypling the Boltzmann equation by the interaction invariant $\psi$ and integrating in the velocity space we get  the \emph{Enskog transport equation} \cite{tesis}:

\begin{equation}
\frac{\partial }{\partial t} [n \langle \psi \rangle]+  \frac{\partial }{\partial x} [n \langle v \psi\rangle]= 0
\label{eq:ete}
\end{equation}
\\
In Eq.(\ref{eq:ete}) the averages must be computed using a consistent ordering in Knudsen's parameter.
To obtain the distribution function up to first order in the Knudsen parameter, Eq.(\ref{eq:o}) is substituted in the Boltzmann equation, so that:

\begin{equation}
\varepsilon f^{(1)}= -\tau ( \frac{\partial f^{(0)}}{\partial t}+ v \frac{\partial f^{(0)}}{\partial x})
\label{eq:x1}
\end{equation}

In general $f=f^{(o)}+\varepsilon f^{(1)}+\varepsilon^2  f^{(2)}+...$, but to first order in the gradients (Navier-Stokes regime), only the first two terms of this expansion are retained. 
\\

Following the CE method, the derivatives present  in Eq.(\ref{eq:x1}) must be calculated using the local equilibrium principle \cite{cc}:
 \begin{equation}
\frac{\partial f^{(0)}}{\partial x}=\frac{\partial f^{(0)}}{\partial n}\frac{\partial n}{\partial x}+\frac{\partial f^{(0)}}{\partial u}\frac{\partial u}{\partial x}+\frac{\partial f^{(0)}}{\partial T}\frac{\partial T}{\partial x}
\end{equation}
and
 \begin{equation}
\frac{\partial f^{(0)}}{\partial t}=\frac{\partial f^{(0)}}{\partial n}\frac{\partial n}{\partial t}+\frac{\partial f^{(0)}}{\partial u}\frac{\partial u}{\partial t}+\frac{\partial f^{(0)}}{\partial T}\frac{\partial T}{\partial t}
\label{eq:x2}
\end{equation}
\\
It is relevant to point out that in the Chapman-Enskog method at the Navier-Stokes level the time derivatives present in Eq.(\ref{eq:x2}) are computed through Eq.(\ref{eq:ete}), \emph{ performing the averages using the equilibrium distribution function} (\ref{eq:mb}). Therefore, $\varepsilon f^{(1)}$ is expressed in terms of thermodynamic forces i.e. spatial gradients of the local equilibrium thermodynamic variables.
\\
\subsection{System of equations in the Euler regime.}

The Euler regime in hydrodynamics is obtained performing the averages of the interaction invariants using $f=f^{(0)}$ :
\\
\\
 for $\psi=1$

\begin{equation}
\frac{\partial n}{\partial t}+ \frac{\partial}{\partial x}(n u)=0
\label{eq:cont}
\end{equation}
\\
Eq. (\ref{eq:cont}) is the well-known \emph{continuity equation}, which expresses particle conservation in the absence of reactions. In order to linearize the transport equations for a static fluid $u_{o}=0$ we shall use:
\begin{equation}
n=n_{0} + \delta n,
\end{equation}
\begin{equation}
u=\delta u 
\end{equation} 
and
\begin{equation}
T=T_{o} + \delta T 
\end{equation} 

keeping terms up to first order in the fluctuations, the linearized version of the  continuity equation becomes : 
\begin{equation}
\frac{\partial}{\partial t} (\delta n)+ n_{0}  \frac{\partial}{\partial x}(\delta u)=0 .
\end{equation}
For the (invariant) linear momentum, we identify  $\psi=mv$,  so that Eq. (\ref{eq:ete}) yields
\begin{equation}
\frac{\partial}{\partial t} (n m \langle v \rangle)+ \frac{\partial}{\partial x}[ n m \langle v^{2} \rangle]=0 
\label{eq:m1}
\end{equation}
Since $<c>=0$  we can rewrite  Eq.(\ref{eq:m1}) as
\begin{equation}
\frac{\partial}{\partial t} (n m u)+ \frac{\partial}{\partial x}[ n m u^{2}]+ \frac{\partial}{\partial x}[ n m \langle c^{2} \rangle]=0 
\label{eq:m2}
\end{equation}
As noted before $\langle c^{2} \rangle =\frac{k T}{m}$, so that

\begin{equation}
\frac{\partial}{\partial t} (nmu)+ \frac{\partial}{\partial x}[ n m u^{2}]+ \frac{\partial}{\partial x}(nkT) =0
\label{eq:m3}
\end{equation}
\\
From classical thermodynamics we can identify the equation of state of  an ideal gas: 
$p=n k T$.
\\ The linearized version of  Eq. (\ref{eq:m3}) in the Euler regime,  using that $\nabla p=n k \nabla T+k T \nabla n $ and that  $ p=p_{0}+\delta p$, reads
\begin{equation}
\frac{\partial}{\partial t} (\delta u)+ \frac{k}{m}\frac{\partial}{\partial x}(\delta T)+ \frac{k T_{0}}{n_{0} m} \frac{\partial}{\partial x}(\delta n) =0
\end{equation}
  \\
\\
We will now consider the energy balance equation.  Using $\psi=\frac{1}{2} m v^{2}$,   Eq. (\ref{eq:ete}) becomes:
\begin{equation}
\frac{\partial}{\partial t}[(n \frac{m}{2} \langle v^{2}\rangle]+ \frac{\partial}{\partial x}[n \frac{1}{2} m \langle v v^{2}\rangle] =0
\end{equation}
Using again that $<c>=0$ it is straightforward to obtain:
 \begin{equation}
\frac{\partial}{\partial t}(\frac{1}{2}mnu^{2})+ \frac{n \partial}{\partial t}(\frac{k T}{2}+\frac{\partial}{\partial x}(\frac{m n u^{2} u }{2})+n u \frac{\partial}{\partial x }(\frac{k T}{2})+\frac{\partial}{\partial x}(n k T u)=0
\end{equation}

The linearized version of  the energy balance equation under Euler's regime becomes:
\begin{equation}
\frac{1}{2} \frac{\partial}{\partial t}(\delta T)+ T_{0} \frac{\partial}{\partial x}(\delta u)=0
\end{equation}

\subsection{Transport equations in the Navier-Stokes regime.}
 
Using the same set of equations but at first order in Knudsen's parameter we have:
\begin{equation}
f= f^{0}+ \varepsilon f^{1}
\end{equation}
\begin{equation}
\langle \psi \rangle =\frac{1}{n}\int f \psi dv 
\end{equation}

In a single component system, the continuity equation (\ref{eq:cont}) remains invariant in both regimes.\\
\\
for $\psi=mv$ equation (\ref{eq:ete}) becomes
\begin{equation}
\frac{\partial}{\partial t}[\int (f^{0}+\varepsilon f^{1})m v d v]+ \frac{\partial}{\partial x}[\int (f^{0}+\varepsilon f^{1})m v v d v ]=0
\label{eq:200}
\end{equation}
the second term yields:
\begin{equation}
\frac{\partial}{\partial x}[n k T+ \int \varepsilon f^{1} mv v dv]=0
\label{eq:28}
\end{equation}
The time derivatives of the equilibrium function present in the first term of Eq.(\ref{eq:200}) are evaluated in the Euler regime so that:
\begin{equation}
\frac{\partial n }{\partial t}= -n \frac{\partial u}{\partial x}-u \frac{\partial n }{\partial x},
\label{eq:juan}
\end{equation}
\begin{equation}
\frac{\partial T }{\partial t}= -\frac{2}{3} T \frac{\partial u}{\partial x}-u \frac{\partial T}{\partial x}
\label{eq:paco}
\end{equation}
and
\begin{equation}
\frac{\partial u }{\partial t}=-\frac{k}{m}\frac{\partial T}{\partial x}-\frac{k T}{n m}\frac{\partial n }{\partial x}-u \frac{\partial u}{\partial x}.
\label{eq:pedro}
\end{equation}
\\
Since $v=(c+u)$, the substitution of  Eqs.(\ref{eq:juan}-\ref{eq:pedro}) in Eq.(\ref{eq:x1}) yields:
\begin{equation}
\varepsilon f^{(1)}=-\tau f^{(0)}[-\frac{3}{2} \frac{c}{T} \frac{\partial T}{\partial x}+\frac{mc^{2}}{2kT^2} c \frac{\partial T}{\partial x}-\frac{2}{3}\frac{\partial u}{\partial x}+\frac{2}{3} \frac{mc^{2}}{ k T}\frac{\partial u}{\partial x}+\frac{m c}{k T}u \frac{\partial u}{\partial x}]
\label{eq:mov}
\end{equation}
substituting  $\varepsilon f^{(1)}$  in Eq. (\ref{eq:28}) we get:         
\begin{equation}
\frac{\partial}{\partial t}(nmu)+\frac{\partial}{\partial x}(nmuu)+\frac{\partial}{\partial x}(nkT)-\frac{\partial}{\partial x}[\eta \frac{\partial u}{\partial x}]=0
\label{eq:29}
\end{equation}
where the shear viscosity coefficient is given by:
\begin{equation}
\eta=\frac{n k T}{6 \sqrt{2}}
\label{eq:eta}
\end{equation}
the linearized version of the momentum balance equation reads:
\begin{equation}
n_{0} m \frac{\partial}{\partial t}(\delta u)+k T_{0}\frac{\partial}{\partial x}(\delta n)+n_{0} k \frac{\partial}{\partial x}(\delta T)-\eta \frac{\partial^{2}}{\partial x^{2}}(\delta u )=0
\label{eq:lin}
\end{equation}
\\
for $\psi=\frac{1}{2}mv^{2}$ the energy balance equation is obtained as follows:
 \begin{equation}
\frac{\partial}{\partial t }[\langle \frac{1}{2} m v^{2}\rangle]+\frac{\partial}{\partial x}[ n \langle \frac{1}{2} m v^{2} v\rangle]=0
\end{equation}
using $v=c+u$, the first term reads:
 \begin{equation}
\frac{\partial}{\partial t }[n\langle \frac{1}{2} m v^{2}\rangle]=\frac{\partial}{\partial t}[\frac{1}{2}m n u^{2}+n\frac{k 
T}{2}]
\end{equation}
the second term yields:
 \begin{equation}
\frac{\partial}{\partial x}[ n \langle \frac{1}{2} m v^{2} v\rangle]=\frac{\partial}{\partial x}[\frac{1}{2}m n u^{2} u+ \frac{nk T u }{2}+n k T u-k_{th} \frac{\partial T}{\partial x}]
\end{equation}
where thermal conductivity $k_{th}$ is given by:
 \begin{equation}
k_{th}=\frac{3}{2}\frac{n k^{2} T \tau}{m}
\label{eq:kappa}
\end{equation}
\\
and the final form of the energy balance equation in its non-linear form reads:
 \begin{equation}
\frac{\partial}{\partial t}(\frac{1}{2}mnu^{2})+n \frac{ \partial}{\partial t}(\frac{k T}{2})+\frac{\partial}{\partial x}(\frac{m n u^{2} u }{2})+n u \frac{\partial}{\partial x }(\frac{k T}{2})+\frac{\partial}{\partial x}(n k T u)- \frac{\partial}{\partial x}(k_{th} \frac{\partial T}{\partial x})=0
\end{equation}
the linear version of the previous equation corresponds to:
 \begin{equation}
\frac{\partial}{\partial t}(\delta T)+2 T_{0} \frac{\partial}{\partial x}(\delta u )-D_{th}\frac{\partial^{2}}{\partial x ^{2}}(\delta T)=0
\end{equation}
where:

 \begin{equation}
D_{th}=\frac{k_{th}}{\frac{1}{2} n_{0} k}=\frac{3 k To}{m} \tau
\label{eq:Dth}
\end{equation}

\newpage

\section{Generic transport equations  up to first order in the  Knudsen parameter.}

 \subsection{Definition of variables}
In this subsection,   the evolution equation for the statistical averages (\ref{eq:ete}) is  applied beyond physical situations.  If the system is taken out of equilibrium and the mean free time is small compared with the characteristic process time, then the counterpart of Eq. (32) is given by:
\begin{equation}
\varepsilon f^{(1)}=-\tau f^{(0)}[-\frac{3}{2} \frac{c}{\Theta} \frac{\partial \Theta}{\partial x}+\frac{mc^{2}}{2k \Theta} c \frac{\partial \Theta}{\partial x}-\frac{2}{3}\frac{\partial u}{\partial x}+\frac{2}{3} \frac{mc^{2}}{ k \Theta}\frac{\partial u}{\partial x}+\frac{m c}{k \Theta}u \frac{\partial u}{\partial x}]
\label{eq:mov2}
\end{equation}
In this equation the  parameter $\Theta$ arises form the Lagrange multipliers method, and it is a measure of the  \textit{variance of the equilibrium distribution function}  yielding
\begin{equation}
\sigma^{2}=\frac{k}{m}\Theta
\end{equation}
In this sense, the parameters $k$ and $m$ do not need an immediate interpretation.

\subsection{ Linearized  Transport equations.}
The linearized system of transport equations previously obtained can be conveniently expressed in terms of the variance of the equilibrium function, namely:

\begin{equation}
\frac{\partial}{\partial t}(\delta n)+n_{0} \frac{\partial}{\partial x}(\delta u)=0
\label{eq:45}
\end{equation}
\begin{equation}
\frac{\partial}{\partial t}(\delta u)+\frac{\sigma_{0}^{2}}{n_{0}}\frac{\partial}{\partial x}(\delta n )+2 \sigma_{0} \frac{\partial}{\partial x}(\delta \sigma)-\sigma_{0}^{2} \tau \frac{\partial^{2}}{\partial x^{2}}(\delta u)=0
\label{eq:46}
\end{equation}
\begin{equation}
\frac{\partial}{\partial t}(\delta \sigma)+ \sigma_{0}\frac{\partial}{\partial x}(\delta u)-3 \sigma_{0}^{2} \tau \frac{\partial^{2}}{\partial x^{2}}(\delta \sigma)=0
\label{eq:47}
\end{equation}
It is appropriate to remark that dissipative effects within the Navier-Stokes regime are taken into account in the last terms in the left hand side of equations (\ref{eq:46}) and  (\ref{eq:47}). In this sense, the transport coefficients (\ref{eq:eta})  and (\ref{eq:kappa}) established on the basis of the BGK approximation can be precisely interpreted in the description of speculative markets.

\subsection{ Analysis of a particular case.}
In this subsection we will consider a simple model of a stock market in which $n$ represents the number of actions in the range of prices ($p$,$p+dp$), $w=\frac{d p}{dt}$ i.e. the price variation in time and $\sigma$ is the variance contained in the equilibrium distribution function. At this point, it is compelling to relate the variable $w$ with the \textit{propensity} introduced in \cite{diez} as a coefficient of the drift term included in the kinetic equation Eq. (2.2) of that work. In this section of our work, the agents correspond to stocks characterized by the price $p$ and its propensity of change $w$, in contrast with kinetic models in which the agents are specified in terms of wealth and propensity to invest. Also, the mean free time first included in Eq. (1) corresponds to the microscopic time in which a single stock varies its price. The characteristic macroscopic time is rather large, since it corresponds (at least) to the length to a full day of work.  Consequently, the Knudsen number $\varepsilon=\frac{\tau}{t_{char}}$ will characterize ideal (Euler) regime for $\varepsilon=0$, a Navier-Stokes regime $0<\varepsilon<<1$ or a rarefied (Burnett) regime $\varepsilon\simeq1$.  A sound statistical approach in which the problem of specifying the mean free time between transactions in markets is given in \cite{once}.
\\
  The system will now read:

\begin{equation}
\frac{\partial}{\partial t}(\delta n)+n_{0} \frac{\partial}{\partial p}(\delta u)=0
\end{equation}
\begin{equation}
\frac{\partial}{\partial t}(\delta u)+\frac{\sigma_{0}^{2}}{n_{0}}\frac{\partial}{\partial p}(\delta n )+2 \sigma_{0} \frac{\partial}{\partial p}(\delta \sigma)-\sigma_{0}^{2} \tau \frac{\partial^{2}}{\partial p^{2}}(\delta u)=0
\end{equation}
\begin{equation}
\frac{\partial}{\partial t}(\delta \sigma)+ \sigma_{0}\frac{\partial}{\partial p}(\delta u)-3 \sigma_{0}^{2} \tau \frac{\partial^{2}}{\partial p^{2}}(\delta \sigma)=0
\end{equation}

where:
\begin{equation}
n(p,t)= \int w f(p,w,t)dw
\end{equation}

$u$ is the average rate in the change of prices:
\begin{equation}
u(p,t)=\langle w \rangle= \frac{1}{n} \int w f(p,w,t)dw
\end{equation}

and
\begin{equation}
\sigma^{2}=\frac{1}{n}\int w^{2} f(p,w,t)dw
\end{equation}

We are now in position to solve the linearized system of equations using the standard procedure of integral transforms. We shall express the Laplace transform (in time) for the fluctuation $\delta X$  by $\delta \tilde{X}$, and the subsequent Fourier transform in the prices domain by  $\delta \hat{\tilde{X}}$,  so that the corresponding algebraic system reads:
\begin{equation}
s \delta \hat{\tilde{n}} +n_{0} \delta  \hat{\tilde{ \theta}} =\delta \tilde{n}(q,0)
\end{equation}
\begin{equation}
-\frac{q^2  \sigma_{0}^{2}}{n_{0}} \delta \hat{\tilde{n}}+(s+q^{2} \sigma_{0}^{2} \tau) \delta \hat{\tilde{\theta}}-2 q^{2} \sigma_{0} \delta \hat{\tilde{\sigma}}=\delta \tilde{\theta}(q,0)
\end{equation}
\begin{equation}
\sigma_{0} \delta \hat{\tilde{ \theta}}+(s+ 3 q^{2}\sigma_{0}^{2} \tau)\delta \hat{\tilde{\sigma}}=\delta \tilde{\sigma} (q,0)
\end{equation}
The dynamics of the system is given by the determinant of the matrix\\ 

\begin{center}

A=
$\left[ 
\begin{array}{ccc}
s & n_{0} & 0 \\ 
\frac{-q^2 \sigma _{0}^{2}}{n_{0}} & s+q^{2}\sigma _{0}^{2}\tau  & 
-2q^{2}\sigma _{0} \\ 
0 & \sigma _{0} & s+3q^{2}\sigma _{0}^{2}\tau 
\end{array}%
\right] $\\
\end{center}

Which reads:
\\\begin{equation}
s^{3}+4 q^{2} \sigma_{0}^{2} \tau  s^{2}+ (3 q^{2} \sigma _{0} ^{2}+ 3 q^{4} \sigma_{0} ^{4}     \tau^{2})s+3 q^{4} \sigma ^{4}_{0} \tau=0
\label{eq:57}
\end{equation}

The solution of the system is expressed in terms of a power series of the mode "q" namely $ s=a_{0}+a_{1} q+ a_{2} q^{2}$. In this approach, terms of order $q^{3}$ and higher are neglected \cite{Mountain}. Thus, the approximated roots of equation  Eq.(\ref{eq:57}) are:
\begin{equation}
s_{1}=- \sigma_{0}^{2} \tau q^{2}
\end{equation}
\begin{equation}
s_{2}=- \frac{3}{2}\sigma_{0}^{2} \tau q^{2}+ \sqrt{3}  \sigma_{0} q i 
\end{equation}
\begin{equation}
s_{3}=- \frac{3}{2}\sigma_{0}^{2} \tau q^{2}- \sqrt{3}  \sigma_{0} q i 
\end{equation}
For the sake of simplicity we will only consider non-vanishing density fluctuations, described by a pulse of height $A$ so that the solution of the system in the Fourier-Laplace space is given by:

\begin{equation}
\delta  \hat{\tilde{\rho}}=A \frac{s^{2}+ 4 q^{2} \sigma_{0}^{2} \tau s}{(s+\sigma_{0}^{2} \tau q^{2}) [(s+\frac{3}{2} \sigma^{2}_{0} \tau q^{2})^{2} +3 \sigma_{0}^{2} q^{2}]}
\end{equation}

The corresponding solution in the $p,t$ domain is obtained using inverse Laplace and Fourier transforms yielding:

\begin{equation}
\delta \rho(p,t)= \frac{A}{2 \sqrt{3 t \tau} \sigma_{0}} e^{-\frac{(p- \sqrt{3} t \sigma_{0})^{2}}{6 t \sigma_{0}^{2} \tau}}
\label{eq:500}
\end{equation}\\

To illustrate this result, we will assign values to the mean free time ($\tau$) and the variance ($\sigma_{0}$) kin to the particular case shown. We will start by imposing a low cost to a set of stocks and a very small mean free time $\tau$ (as required by the CE method) to see how the system evolves in time.  Substituting these values in Eq.(\ref{eq:500}), the solution yields a behavior analog to the one known for diffusion (see figure 1). \\  \\
\newpage
\begin{figure}
\caption{Evolution of density fluctuations in arbitrary units in the p-t space. This plot corresponds to $\sigma_{0}=0.5$ and $\tau=0.05$  and resembles a "pulse" in which the fluctuation of the density of prices decreases as the price and time increase. }
\includegraphics[
  width=6in,
  height=4in]
{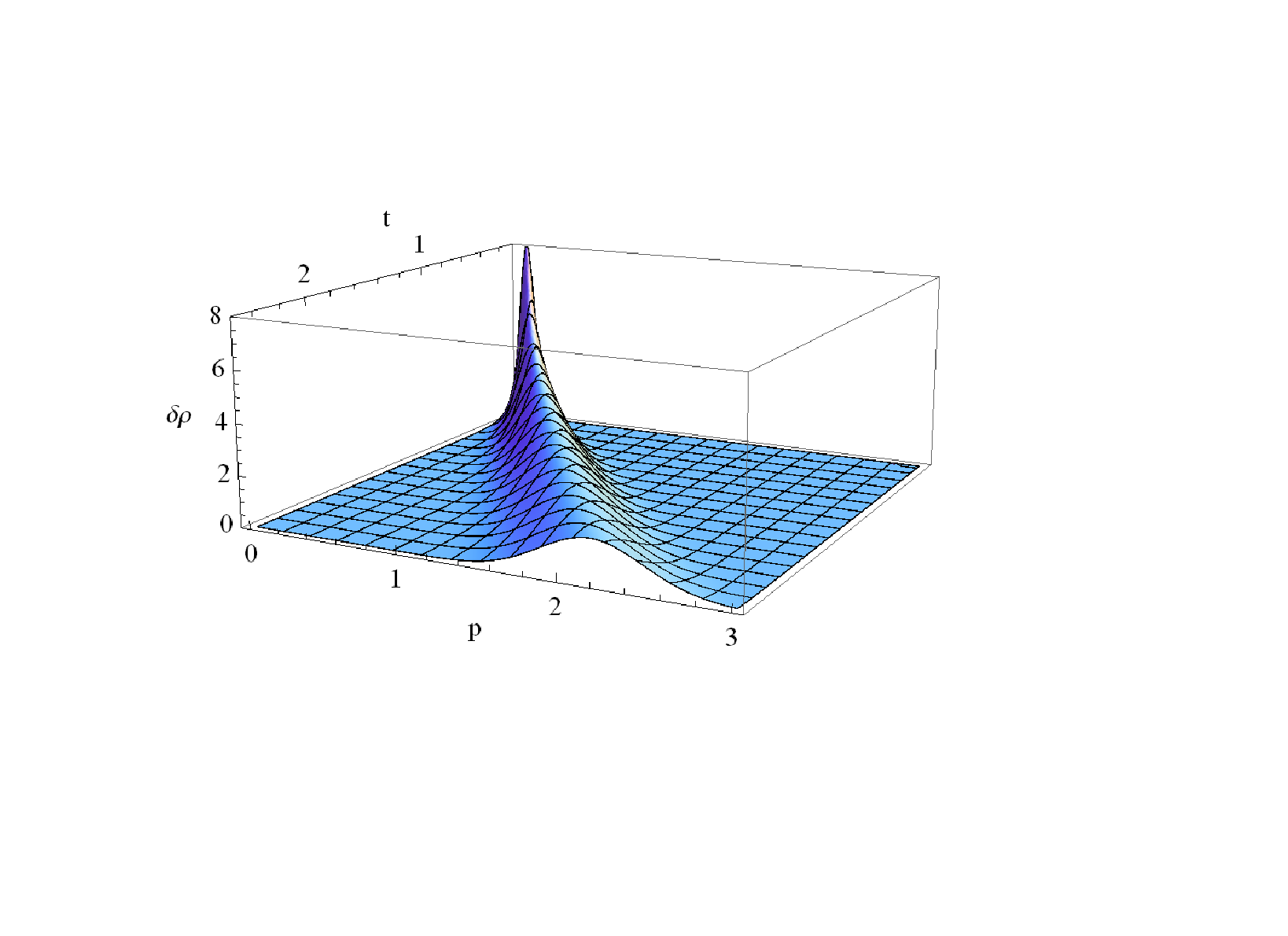}
\label{fig:fig1}
\end{figure}

\newpage

\section{Final remarks}

The main result of this work corresponds to the generalized, linearized system of equations (46-48), which is valid to describe the evolution of fluctuations present in any system describable in terms of a distribution function, a maximum entropy principle and a Knudsen parameter lower than unity. This approach is different from similar formalisms based on the Boltzmann equation in which the collisional kernel is modeled using a Fokker-Planck approximation [10] [12]. In our work, the use of a BGK type-kernel yields a complete set of transport equations free of adjustable parameters. 
Our proposal seems to be applicable to a broad spectrum of fields. For example, a financial market in which the number of stocks within a range of prices is describable by equations (46-48) will exhibit a behavior analog to the one present in hydrodynamics. Features such as instabilities, turbulence, cyclic properties, pattern formation and self-organization will arise from the non-linear counterpart of equations (46-48), allowing a deeper understanding of complex phenomena present in economic processes. This kind of features will also be exportable to any other situation which complies with the assumptions involved in the Chapman-Enskog method for solving the master equation. Moreover, the proper definition of the Knudsen number as an ordering parameter allows to identify rarefied regimes in speculative markets in which the analog of the Burnett regime might be applicable to the description of short periods of time in trading hours 
 \cite{trece}.  \\

The authors wish to thank Dominique Brun-Battistini for providing helpful comments for the final version of this work.


\begin{thebibliography}{10}

\bibitem[1]{Jaynes} E.T. Jaynes,  Phys. Rev.  \textbf{44}, 139-145 (1957). 
\bibitem[2]{Shannon}  C.E Shannon, Bell System Tech. J. \textbf{27}, 379, 623 (1948).

\bibitem[3]{cc} S. Chapman and T.G Cowling, "The  mathematical theory of non-uniform gases", third edition,  Cambridge University Press  (1970). 

\bibitem[4]{BGK}  P.L. Bhatnagar, E.P. Gross, M. Krook, Phys. Rev.  \textbf{94},  511–525 (1954).

\bibitem[5] {CC} A. S. Chakrabarti, B. K. Chakraborti. "Microeconomics of the ideal gas like market models". Physica A 388: 4151–4158 (2009).
\bibitem[6] {tesis} J. Nassios, "Kinetic Theory an the BGK equation: Gas Dynamics for the Nanoscale".Honours thesis (2008).
\bibitem[7]{bernepecora}B. J. Berne \& R. Pecora: \textquotedblright{}Dynamic
light scattering with applications to chemistry, biology and physics\textquotedblright{},
Dover Publ. NY (2000).
\bibitem[8] {CC} L. S. Garcia-Colin and P.Goldstein," Procesos Irreversibles:Teoria y Aplicaciones", El Colegio Nacional-UAMI, Mexico (2013). In Spanish. 
\bibitem[9] {Mountain} Mountain,R.D.," Spectral distribution of scattered light ina simple fluid", Rev. Mod. Phys. 38,205, (1966).
\bibitem[10] {diez} During, B. Toscani, G.," Hydrodynamics from kinetic models of conservative economies", Physica A, 384,(2007),493-506.
\bibitem[11] {once} Y. Wang, N. Ding, L. Zhang. "The circulation of money and holding time distribution." Physica A 324(3-4), 665-667 (2003).
\bibitem[12] {doce}Giovanni Naldi, Lorenzo Pareschi, Giuseppe Toscani, "Mathematical Modeling of Collective Behavior in Socio-Economic and Life Sciences", Birkhauser, Boston (2010).
\bibitem[13] {trece} L.S. Garcia-Colin, R. M. Velasco and F. J. Uribe, "Beyond the Navier-Stokes equations: Burnett hydrodynamics", Phys. Rep. 465, 149-189 (2008).
\end{thebibliography}
\end{document}